\documentclass{aa}

\usepackage{graphics}

\usepackage{journal}

\usepackage{aabib99}

\begin{document}

\thesaurus{11(11.05.2; 13.21.1; 13.09.1; 13.18.1)}

\title{Galaxy Modelling --- II. Multi--Wavelength Faint Counts from
a Semi--Analytic Model of Galaxy Formation}

\author{ Julien E.\,G.\,Devriendt\inst{1,2}
         \and Bruno Guiderdoni\inst{1}} 

\institute{ Institut d'Astrophysique de Paris,
            98\,{\it bis} Boulevard Arago, F--75014 Paris, France
            \and Nuclear \& Astrophysics Laboratory,
            Keble Road, Oxford OX1 3RH, United Kingdom} 

\offprints{\tt jeg@astro.ox.ac.uk} 

\date{Received ??  / Accepted ??}

\authorrunning{J. Devriendt \& B. Guiderdoni}

\titlerunning{II. Multi--Wavelength Counts from a Semi--Analytic Model}

\maketitle

\begin{abstract}

This paper predicts self--consistent
faint galaxy counts from the UV to the submm wavelength range.
The {\sc stardust} spectral energy distributions described 
in Devriendt et al. \cite*{DGS99} (Paper I) are embedded within the 
explicit cosmological framework of a simple semi--analytic model of galaxy 
formation and evolution. We begin with a description of the non--dissipative 
and dissipative collapses of primordial perturbations, and  
plug in standard recipes for star formation, stellar
evolution and feedback. We also model the
absorption of starlight by dust and its re--processing in the IR and submm. 
We then build a class of models which capture the luminosity 
budget of the universe through faint galaxy counts and redshift
distributions in the whole wavelength range spanned by our spectra.
In contrast with a rather stable behaviour in the optical and even in 
the far--IR, the submm counts are dramatically sensitive to
variations in the cosmological parameters and changes in the star formation 
history.
Faint submm counts are more easily accommodated within an open universe
with a low value of $\Omega_0$, or a flat universe with a non--zero 
cosmological constant. We confirm the suggestion of Guiderdoni et al. 
\cite*{GHBM98} that matching the current multi--wavelength data requires 
a population of heavily--extinguished, massive galaxies 
with large star formation rates ($\sim 500~M_\odot$ yr$^{-1}$)
at intermediate and high redshift ($z \geq 1.5$).
Such a population of objects probably is the consequence of an 
increase of interaction and merging activity at high redshift, but a 
realistic quantitative description can only be obtained through  
more detailed modelling of such processes.
This study illustrates the implementation of multi-wavelength 
spectra into a semi--analytic model. In spite of its simplicity, 
it already provides fair fits of the current data of faint counts,
and a physically motivated way of interpolating and extrapolating these data 
to other wavelengths and fainter flux levels.


\end{abstract}

\section{Introduction}

In the dense gas clouds that harbour starbursts, the ultraviolet (UV) 
light of young stars is absorbed by dust grains 
which, in turn, release their thermal energy at infrared (IR) 
and submillimetre (submm) wavelengths. Thus,
understanding the star formation history of galaxies clearly requires a 
correct assessment of the UV to submm luminosity budget. The most 
straightforward and simple observational probe of such a luminosity budget 
is the analysis of the faint galaxy counts obtained at various wavelengths. 
In this purview, this paper proposes self--consistent theoretical 
predictions of faint galaxy counts at optical, 
IR and submm wavelengths that can be directly compared with the current 
host of data, and used to prepare observational strategies with forthcoming 
instruments.

In the local universe, only 
30 \% of the bolometric luminosity is released in the IR/submm wavelength 
range \cite{SN91}, and the effect of extinction can thus be 
considered as a mere correction
that does not change the main evolutionary trends elaborated from optical 
studies. However, there is now a growing amount of evidence that this 
fraction was much higher in the past. Indeed,
the discovery of the Cosmic Infrared Background (CIRB) 
at a level ten times higher 
than the no--evolution predictions based on the {\sc iras} local IR luminosity 
function, and twice as high as the Cosmic Optical Background obtained from 
optical counts, showed that dust extinction and emission play a major 
role in defining the luminosity budget of high--redshift galaxies
(Puget et al. \cite*{P+al96}, Guiderdoni et al. \cite*{GBPLH97}, 
Schlegel et al. \cite*{SFD98}, 
Fixsen et al. \cite*{FDMBS98}, Hauser et al. \cite*{H+al98}). Since this 
major breakthrough, deep surveys with the {\sc iso} satellite at 15 $\mu$m
(Oliver et al. \cite*{O+al97}, Aussel et al. \cite*{ACES99}, 
Elbaz et al. \cite*{EC+al99}) and 
175 $\mu$m (Kawara et al. \cite*{K+al98}, Puget et al. \cite*{P+al99}),
and with the SCUBA instrument at 850 $\mu$m (Smail et al. \cite*{SIB97},
Barger et al. \cite*{BCSFTSKO98}, Hughes et al. \cite*{HS+al98}, Eales
et al. \cite*{ELGDBHLC99}, Barger et al. \cite*{BCS99}) have begun to 
break the CIRB into its brightest contributors. Although identification and 
spectroscopic follow--up of the submm sources are not easy, the preliminary 
results of such studies seem to show that part of these sources are 
the high--redshift counterparts of the local luminous and 
ultraluminous IR galaxies (LIRGs and ULIRGs) discovered by {\sc iras}
(Smail et al. \cite*{SIBK98}, Lilly et al. \cite*{L+al99}, 
Barger et al. \cite*{BCSIBK99}). 
In parallel to this pioneering exploration of the ``optically--dark''
and ``infrared--bright'' side of the
universe, a more careful examination of
the Canada--France Redshift Survey (CFRS) galaxies at $z\sim 1$, 
and Lyman Break Galaxies at $z \sim 3$ 
and 4 do show a significant amount of extinction (Flores et al. 
\cite*{FH+al99}, Steidel et al. \cite*{SAGDP99}, Meurer et al. 
\cite*{MHC99}). The previous estimates of 
the UV fluxes, and consequently of the star formation rates, in these 
objects have to be respectively multiplied by factors 3 and 5 
to take into account the effect of extinction. Dust seems to be present at 
still higher redshifts. For instance, it is seen in a lensed galaxy 
at $z=4.92$ \cite{SNFMI98} and even in a Lyman $\alpha$ galaxy at 
$z=6.68$ \cite{CLP99}.

The synthetic spectra of stellar populations in galaxies are easily computed 
from spectrophotometric models of galaxy evolution. Unfortunately, 
most of these models neglect the influence of dust on the spectral appearance 
of galaxies. Guiderdoni \& Rocca--Volmerange \cite*{GRV87} proposed a 
first modelling of the effect of dust extinction. Later, Mazzei et 
al. \cite*{MXZ92} basically used the same recipe for extinction, but they also
computed dust emission to get spectral energy distributions (SEDs) from the 
UV to the far--IR. Complete sets of synthetic spectra are now available 
from the {\sc grasil} \cite{SGBD98} and {\sc stardust} models 
(Devriendt et al. \cite*{DGS99}, hereafter Paper I) of spectrophotometric 
evolution. These models share the same spirit, but they differ by a number 
of details. The {\sc grasil} SEDs in the IR are computed from a more 
sophisticated model of transfer, that is more explicit physically, 
but involves several free parameters, whereas {\sc stardust} SEDs in the IR 
are computed with a minimal number of free parameters, by weighing various 
dust components to reproduce the observed {\sc iras} colour--luminosity 
relations. 

These spectra can be used in phenomenological models of faint galaxy counts, 
that extrapolate the evolution of the local galaxies backwards under the 
assumption of pure luminosity evolution. For instance,
the predictions of faint galaxy counts at optical wavelengths by 
Guiderdoni \& Rocca--Volmerange \cite*{GRV90} used their optical spectra with 
extinction, whereas Franceschini et al. \cite*{FZTMD91}, 
\cite*{FMZD94} used the Mazzei et al. spectra with dust extinction 
and emission to produce the first set of counts at optical {\it and} FIR 
wavelengths. However, semi--analytic models of galaxy formation 
(hereafter SAMs) are a much 
more powerful approach to describe the physical processes that rule 
galaxy formation and evolution within an explicit cosmological context
(White \& Frenk \cite*{WF91}, Lacey \& Silk \cite*{LS91}, 
Kauffmann et al. \cite*{KWG93}, Cole et al. \cite*{CAFNZ94}, 
Somerville \& Primack \cite*{SP00}). 
White \& Frenk \cite*{WF91}, Lacey et al. \cite*{LGRS93}, 
Kauffmann et al. \cite*{KGW94}, and Cole et al. \cite*{CAFNZ94}
proposed predictions of faint galaxy counts at optical wavelengths 
(basically the $B$ and $K$ bands) from their models. However, predictions at 
IR/submm wavelengths from a SAM were produced much later by Guiderdoni et al. 
(\cite*{GBPLH97}, \cite*{GHBM98}, hereafter GHBM).
But this first study did not give the corresponding predictions at optical 
wavelengths, and was restricted to the $\Omega_0=1$ standard Cold Dark 
Matter (CDM) model. 

In this paper, we implement the {\sc stardust} spectra into a SAM 
to make predictions of faint galaxy counts and redshift distributions from
the UV to the submm wavelength range, very much in the spirit of GHBM. 
We also extend the SAM to other cosmologies, and study the sensitivity of 
the results to the cosmological parameters and star formation history. 
Although our approach has a number of shortcomings which are due to the 
simplicity of our model, this paper primarily 
intends to show that (i) the implementation of {\sc stardust} SEDs into SAMs 
is straightforward because of its small number of 
free parameters;~ (ii) the model gives fits that  
are already very satisfactory in spite of the simplicity of the approach;~
and (iii) the outputs are a physically motivated tool to interpolate or 
extrapolate the current observations of faint counts to other wavelengths 
and/or flux levels. 
This is needed to prepare the observational strategies with the 
forthcoming IR/submm satellites {\sc sirtf}, {\sc first} and {\sc planck}, 
as well as with the Atacama Large Millimetre Array. 

In section 2, we briefly describe how we connect the {\sc stardust} spectra
with the various physical processes that are relevant to galaxy
formation, within our SAM. We point out the differences with GHBM.
Section 3 discusses the values of the free parameters that define the 
so--called ``quiescent'' mode of star formation.
Section 4 focuses on the sensitivity of the faint galaxy counts
to a change in the cosmological parameters.
Section 5 studies the sensitivity of the faint galaxy counts to 
galaxy evolution, and more specifically to the presence of a 
heavily--extinguished ``starburst'' mode of star formation similar to the one 
in ULIRGs. Finally, a fiducial model is proposed. We discuss our results in 
Section 6.

\section{The basics of our semi--analytic model}
 
In the SAM {\em ab initio} approach, galaxies form from Gaussian 
random density fluctuations in the primordial matter distribution, 
dominated by CDM.
Bound perturbations grow along with the expanding universe, until
gravitation makes them turn around and (non--dissipatively) collapse.
As a result, they end up as virialized halos. Then the 
collisionally--shocked baryonic gas cools down radiatively, and 
settles at the bottoms of the potential wells where it is 
rotationally--supported. Stars form from the cold gas
and evolve. At the end of their lifetimes, they inject energy, gas and heavy 
elements back into the interstellar medium. The chemical evolution is 
computed, and the recipes developed in GHBM and in Paper I give the amount 
of optical luminosity that is absorbed by dust and thermally released at 
IR/submm wavelengths. Finally, overall SEDs 
from the UV to the submm are computed with {\sc stardust}. 

We refer the reader to GHBM for a detailed description of how to 
compute the mass distribution of collapsed dark matter halos from the 
peaks formalism introduced by Bardeen et al. \cite*{BBKS86}, and
Lacey and Silk \cite*{LS91} in an Einstein--de Sitter universe.
We give in appendix A the quantities
which enable us to extend this formalism to low matter--density 
universes with or without a cosmological constant. 
As we follow closely the prescriptions in GHBM, we only mention in
the following subsections the quantities which differ from their work.

\subsection{Gas}

We assume that a universal ``baryonic fraction'' 
$\Omega_B/\Omega_0$ of the pristine gas gets locked up within each dark matter 
halo, where it is collisionally ionised by the shocks occurring during 
virialization. Because it can cool radiatively, gas then sinks into the 
potential wells of the halos. 
The cooling time depends on the gas metallicity.
Here, we decide to adopt the cooling function given by
Sutherland and Dopita \cite*{SD93} for one third of solar metallicity.
This choice is motivated by the fact that
it is the average value that is observed today in clusters, and 
probably, as argued by Renzini \cite*{R99}, the average value
of the low--redshift universe as a whole. Under this assumption, the 
cooling time is underestimated in high--redshift halos where the gas is
more metal--poor. However, these objects are also smaller and denser on
an average, so that their cooling times are already very short.

We assume that the gas stops falling into the
dark matter potential wells when it reaches rotational equilibrium, and
forms rotating thin disks (see e.g. Dalcanton et al. \cite*{DSS97} and
Mo et al. \cite*{MMW98a}).
Following these authors, we adopt for the thin disk an exponential surface 
density profile with scale length $r_d$, and truncation radius $r_t$, 
such as:
\begin{displaymath} 
\Sigma(r) = 
\left\{
\begin{array}{ll} 
\Sigma(0) \exp(-\frac{r}{r_d})& \mbox{if $ r \le r_t $ ,} \\
0& \mbox{if $ r > r_t $}
\end{array} \right.  
\end{displaymath}
where $r_t$ is defined as the minimum
value between the virial radius $r_{vir}$, and $f_c r_d$.  The free parameter
$f_c$ defines the extent of the cold gas disk.

We then relate the exponential scale length of the cold
gas disk $r_d$ to the initial radius $r_{vir}$, through conservation of 
specific angular momentum \cite{FE80}. 
As shown by Mo et al. \cite*{MMW98a}, stability criteria yield:
\begin{eqnarray}
r_d = \frac{1}{\sqrt{2}} \lambda  r_{vir} \times \left( 1 - \frac{f_c^2
  \exp(-f_c)}{2 \, [1 - (1+f_c) \exp(-f_c)]} \right)^{-1} ,
\label{rd}
\end{eqnarray}
where $\lambda \equiv J|E|^{1/2}G^{-1}M^{-5/2}\simeq 0.05 \pm 0.03$
is the well--know dimensionless spin parameter. 

As this will be important later, we emphasise that the simple formalism 
used here does not allow us to form 
spheroids through mergers/interactions of galaxies.
Therefore, in section 5, we will define a ``starburst mode''
which phenomenologically accounts for this process. 
 
\subsection{Stars}

The only time scale available in our gas disks is the dynamical 
time scale $t_{dyn} \equiv 2\pi r_d /V_c$.
Therefore, guided by observational data (Kennicutt 
\cite*{K98}), we assume that the complicated physical processes ruling
star formation lead, at least in a disk galaxy,
to a global star formation rate (SFR) with the simple law:
\begin{eqnarray} 
\mathrm{SFR}(t) = \frac{M_{gas}(t)}{\beta t_{dyn}} . 
\end{eqnarray}
where $M_{gas}(t)$ is the total mass of cold gas in the disk at time $t$. 
We introduce an efficiency factor $\beta^{-1}$ as a second free parameter.
The IMF is chosen to be Salpeter's, with slope $x=1.35$ between masses
$m_d=0.1$ and $m_u=120 M_\odot$. 

The {\sc stardust} spectrophotometric and chemical evolution model 
presented in Paper I is then used to compute metal enrichment of the gas as 
well as the UV to NIR spectra of the stellar populations produced with such 
star formation rates. Details on the stellar spectra, evolutionary tracks and 
yields can be retrieved from this paper and references therein.

\subsection{Feedback}

Along with producing metals, massive stars which, at the end of 
their lifetimes,
explode in galaxies, eject hot gas and heavy elements 
into the interstellar and/or intergalactic medium. We focus here on    
the modelling of this ``stellar feedback'', which is inspired from 
Dekel and Silk \cite*{DS86}.
The average binding energy of a mass of gas  
$M_{gas}(t)$ distributed within a truncated exponential disk at time $t$,
which is gravitationally--dominated by its dark matter halo, is given by:
\begin{eqnarray}
\frac{1}{\pi r_t^2}\int_0^{r_t} 2 \pi r M_{gas}(r,t) \Phi (r) \mathrm{d}r & \simeq & 
 \frac{1}{2} M_{gas}(t) \left[ V_c^2 + V_{esc}^2 (r_t) \right] , 
\end{eqnarray}
where $\Phi (r)$ is the gravitational potential of
a singular isothermal sphere truncated at virial radius $r_{vir}$, 
and:
\begin{eqnarray}
V_{esc} (r) = V_c \left[2 \, \left(1-\ln \frac{r}{r_{vir}} 
\right) \right]^{1/2} 
\end{eqnarray}
is its escape velocity at radius $r$.

As a result, the energy balance between the gravitational
binding energy and the kinetic energy pumped by supernovae into 
the interstellar medium yields the fraction of stars $F_\star$ that 
formed {\em before} the triggering of the galactic wind (at time $t_W$):
\begin{equation}
F_\star = \frac{M_\star(t_W)}{M_\star(t_W) + M_{gas}(t_W)} = 
\frac{1}{1 + (V_{hot}/V_c)^2} ,
\label{fst}
\end{equation}
with:
\begin{eqnarray}
V_{hot} \equiv \left[ \frac{2\eta_{SN}E_{SN}\epsilon_{SN}}{1 + (V_{esc}
    (r_t)/V_c)^2} \right]^{1/2} ,
\end{eqnarray}
where the energy available per supernova is $E_{SN} = 10^{51} \mathrm{erg}$, 
and the number of supernovae per mass unit of the stars that just formed is 
$\eta_{SN}=7.5 \times 10^{-3}$ for a Salpeter IMF. 
The SN heating efficiency $\epsilon_{SN}$ is a third free parameter. 
By taking the initial gas mass available for star formation to be 
the initial cold gas mass minus the gas mass lost in the galactic
wind, one then approximate chemical evolution by the closed--box model 
described in Paper I. Note that we have neglected any dynamical effect 
due to mass loss in the previous analysis. 
 
\subsection{Dust}

Part of the luminosity released by stars is absorbed by dust and re--emitted in
the IR/submm range. We now briefly outline how we 
compute the luminosity budget of our objects within the SAM. We emphasise
that this is a major improvement with respect to GHBM, as for the first time,
stellar and dust emission are linked self--consistently. 
As in Paper I, we proceed to derive the IR/submm dust spectra with 
three steps:
(i) computation of the optical depth of the disks, (ii)
computation of the amount of bolometric energy absorbed by dust, and (iii)
computation of the spectral energy distribution of dust emission.  

The first step is easily completed because we know the sizes 
of our objects from eq.~\ref{rd} and the definition of the truncation 
radius $r_t$, and we obtain the mass of gas 
$M_{gas}(t)$ and metallicity $Z_g(t)$ as a function of time through our 
model of chemical evolution. We then use the scaling 
of the extinction curve with gas column density and metallicity 
described in Guiderdoni \& Rocca-Volmerange \cite*{GRV87} to compute
the face--on optical depth of our objects at any wavelength:
\begin{equation}
\tau_\lambda^\mathrm{z} (t)  = \left( {A_\lambda \over
A_\mathrm{V}} \right)_{Z_\odot} \left( {Z_g(t) \over Z_\odot} \right)^s 
\left({\langle N_{\sc\mathrm H}(t) \rangle \over
2.1 \times 10^{21} {\mathrm~at~cm^{-2}}} \right) \, ,
\end{equation} 
where the mean H column density (accounting for the presence of
helium) reads:
\begin{eqnarray}
\langle N_\mathrm{H}(t) \rangle = {M_{gas}(t) \over 1.4 \mu m_p \pi r_t^2} \, .
\end{eqnarray} 
 
\begin{figure}[htbp]
\resizebox{9cm}{!}{\includegraphics{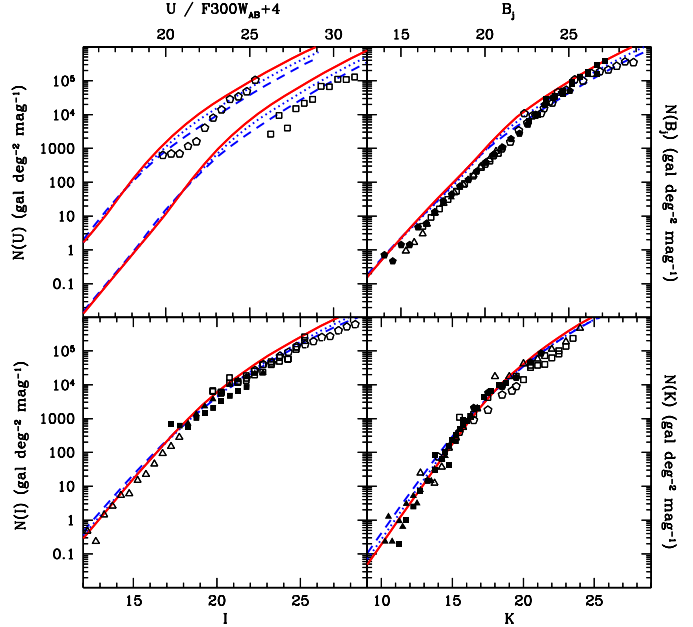}}
\hfill
\caption{Influence of the different 
cosmologies on the UV/near--IR faint counts.
Dots stand for $\Lambda$CDM, dashes for OCDM,
and solid lines for SCDM.  
Data are from Hogg et al. (1997) (U band), 
Williams et al. (1996) (F300W$_\mathrm{AB}$, B \& I bands), 
Arnouts et al. (1997) (B band), Bertin \& Dennefeld (1997) 
(B band), Gardner et al. (1996) (B, I \& K bands), Metcalfe et al. (1995)
(B band), Weir et al. (1995) (B band), Smail et al. (1995) (I band), 
Le F\`evre et al. (1995) (I band), 
Moustakas et al. (1997) (K band), and Djogorvski et al. (1995) (K band).}
\label{ccopt}
\end{figure}

\begin{figure}[htbp]
\resizebox{9cm}{!}{\includegraphics{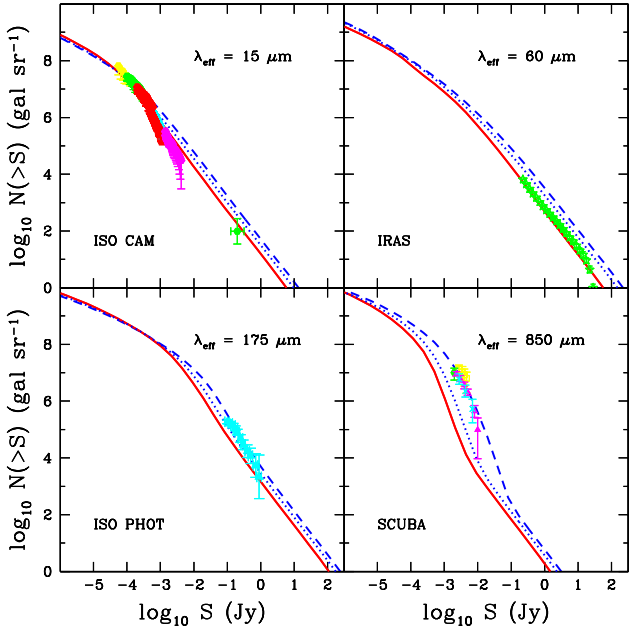}}
\hfill
\caption[]{Influence of the different 
cosmologies on the mid--IR/FIR/submm counts. 
Data are from Elbaz et al. (1999) (15 $\mu$m),
Kawara et al. (1998) and Puget et al. (1999)
(175 $\mu$m), Smail et al. (1997), Eales et al. (1999) and Barger et al. 
(1999a) (850$\mu$m).
Coding for the lines is the same as in Fig~\ref{ccopt}.}
\label{ccinf}
\end{figure}

The second step is more delicate because it involves choosing a ``realistic'' 
geometry distribution for the relative distribution of stars and dust.
We model galaxies as oblate ellipsoids where dust and stars are 
homogeneously mixed, and scattering is taken into account. As explained in 
Paper I, the model gives a decent fit of the sample of local 
spirals analysed by Andreani \& Franceschini \cite*{AF96}.

Finally, the third step involves an explicit modelling of the dust 
grain properties and sizes.  We use the three--component model 
described in D\'esert et al. \cite*{DBP90} for the Milky Way 
with polycyclic aromatic hydrocarbons,
very small grains and big grains, and we allow a fraction of the 
big grain population to be in thermal
equilibrium at a warmer temperature if our galaxies undergo a massive 
starburst. The weights of these four components are fixed in order to 
reproduce the relations of IR/submm colours with bolometric 
IR luminosity $L_\mathrm{IR}$ that are observed locally, 
as detailed in Paper I. Once
the full (UV/submm) spectral energy distributions of individual
objects are computed following such a method, we build populations of galaxies 
for which we derive galaxy counts and redshift distributions. 
We present these results in the following sections.

\section{The free parameters}

In addition to the cosmological parameters $h$, $\Omega_0$, $\lambda_0$, 
$\Omega_B$ and $\sigma_{8h^{-1}}$, and to the choice of the IMF, 
which is assumed to be 
constant throughout a Hubble time, we basically have three astrophysical free 
parameters in the current version of our simple semi--analytic model:~ 
the star formation efficiency $\beta^{-1}$, the SN heating  
efficiency $\epsilon_{SN}$, and the disk truncation parameter $f_c$ which 
is used to compute the gas column density and the face--on optical depth. 
As a matter of fact, there is not much freedom in the choice of these 
parameters.

First, the value of the star formation efficiency $\beta^{-1}$ 
deduced from Kennicutt's data \cite{K98} is about $\beta \simeq 50$ for 
our definition, and is valid for galaxies ranging from quiescent 
objects to very active starbursts. We refer to GHBM for a discussion of how 
this prescription actually compares to the data, especially to the so--called 
``Roberts times'' in nearby disks, and just mention that the 
difference between the value in this paper and the one used in GHBM
($\beta= 100$) stems from the different prescriptions used
to compute disk sizes, which result in our disks being about
twice as large as theirs.
As mentioned by Kennicutt, there is a lot of scatter in the data 
($\pm$ 30--50 \%), which, along with plausible systematics in the 
calibration of the different star formation estimators, should make the value 
of $\beta$ uncertain by at least 20 \%. Increasing $\beta$ decreases the 
normalisation and slope of the optical and IR counts, because
star formation is lower, and takes place at lower redshifts.

Second, recent numerical simulations \cite{TGJS98} suggest that the SN 
heating efficiency $\epsilon_{SN}$ is $\simeq 0.09$. However, there is much 
uncertainty on the actual efficiency of SN explosions in a disk galaxy, 
because SN bubbles can blow their energy out of the disk without altering
the cold gas (see e.g. De Young \& Heckman \cite*{YH94}, and Lobo \& Guiderdoni
\cite*{LG99} for an examination of the issue within a SAM). Consequently, 
$\epsilon_{SN}$ could be very low. We adopt $\epsilon_{SN} = 0.03$ in the following.
Increasing $\epsilon_{SN}$ decreases the 
normalisation and slope of the optical and IR counts, since star formation is 
quenched in galaxies with still higher masses, that form at still lower 
redshifts.

Third, the average value of the disk truncation parameter $f_c$, 
which measures the gaseous disk extension, is around 6, from the  
sample of spiral galaxies with various morphological types observed by 
Bosma \cite*{B81}, and used by GHMB. However, this number is probably 
uncertain by about a factor 2, and it can be adjusted within this range 
in order to match the UV/optical/near--IR counts as well as {\sc iras} 
counts, as far as this parameter fixes the amount of dust absorption in a 
disk galaxy. Increasing $f_c$ increases the 
normalisation of the optical counts and decreases that of the IR counts, 
since extinction decreases.

As explained in Paper I, the set of {\sc stardust} spectra depends on (i) 
the mass of baryons in the galaxy, 
(ii) the star formation timescale $t_*$, (iii) the age $t$ 
of the stellar population, and
(iv) the parameter called $f_H$ that links the gas mass fraction to the gas 
surface density, and is used in the computation of the face--on optical depth. 
In the SAMs, these quantities are
computed directly from the cold gas mass $M_{bar}$, the dynamical time 
$t_{dyn}$, the collapse redshift $z_{coll}$ and observed redshift $z$, 
and the disk exponential length $r_d$, provided the values of the 
cosmological parameters are chosen, and the astrophysical parameters $\beta$, 
$\epsilon_{SN}$ and $f_c$ are fixed. Thus the implementation of the
{\sc stardust} spectral energy distributions within our SAM is very 
straightforward and does not bring new free parameters.

The above--mentioned values of the free parameters define the so--called 
``quiescent mode'' of star formation, similar to what is observed in local 
disks. As a reference point, we list, for the $\Lambda$CDM cosmology,  
the properties of a disk galaxy hosted by a halo of about 
$7.5 \times 10^{11} M_\odot$, 
which collapses at a redshift $\sim 3$ (meaning that 
the age of the ``Milky Way''--class spiral galaxy that sits in this halo
is about 11.4 Gyr) with a spin parameter $\simeq 0.08$. 
At redshift 0, such a galaxy has turned about 88 \% of
its total 7.5 $\times 10^{10} M_\odot$ of cold gas into stars.
Its disk exponential scale length is about 3.5 kpc, yielding a 
gaseous disk extending to 21 kpc with an average hydrogen column
of $8 \times 10^{20} \mathrm{at~cm^{-2}}$ and a metallicity of 0.02. 
This, in turn implies a face on optical depth in the {\em B} band of 0.7
resulting in values of $M_B = -19.8$ and $L_\mathrm{IR} 
= 2 \times 10^{10} L_\odot$ 
for the face--on absolute $B$ magnitude and the bolometric IR luminosity
(between 3 and 1000 $\mu$m) respectively.

Although this discussion is only valid, strictly speaking, for a
given cosmology, it is unlikely that different cosmological parameters will 
significantly affect physical parameters like star formation or feedback
efficiency. We therefore consider our astrophysical parameters 
as independent of the cosmological model. In the next section, we 
keep the same values of the astrophysical parameters, and we 
study the predictions of the SAM for the sets of cosmological parameters 
that are displayed in table~\ref{t_param}.

\begin{figure}[htbp]
\resizebox{9cm}{!}{\includegraphics{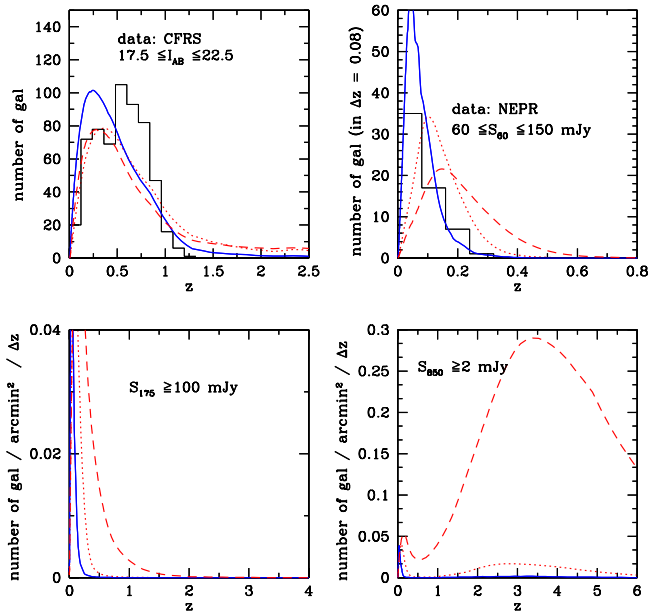}}
\hfill
\caption{Influence of the different 
cosmologies on multi--wavelength redshift distributions
of galaxies. Coding for the lines is the same 
as in Fig~\ref{ccopt}. Data are from Crampton et al. (1995) 
(Canada--France Redshift Survey), and Ashby et al. (1996) 
(North Ecliptic Pole region). The predicted curves in the $I$ band 
and at 60 $\mu$m have been renormalised to the total number of galaxies 
in the data.}
\label{rdec}
\end{figure}

\section{Sensitivity of faint counts to cosmological parameters}

We emphasise that our purpose here is not to determine values of $h$,
$\Omega_0$ or $\lambda_0$, but rather to answer the 
following question: what is the net effect of changing the cosmological 
parameters on faint galaxy counts from the UV to the submm. 
We explore the set of cosmological parameters displayed in 
table~\ref{t_param}. All the astrophysical parameters are fixed at the 
``natural values'' of the ``quiescent'' mode of star formation based on the
local universe.

\begin{table}[t]
\centerline{
\begin{tabular}{|p{2.8cm}||p{0.4cm}|p{0.4cm}|p{0.4cm}|p{1.2cm}|p{0.7cm}|}
\hline
Cosmological Models & $\Omega_0$ & $\lambda_0$ & $h$ & $\Omega_B$ &
$\sigma_{8h^{-1}}$ \\
\hline
 & & & & &  \\ 
SCDM         & 1.0 & 0.0 & 0.5 & 0.015$\,h^{-2}$ & 0.58 \\
$\Lambda$CDM & 0.3 & 0.7 & 0.7 & 0.015$\,h^{-2}$ & 1.0 \\
OCDM         & 0.3 & 0.0 & 0.7 & 0.015$\,h^{-2}$ & 1.0 \\
 & & & & &  \\ 
\hline
\end{tabular}}
\caption []{Parameters of the different cosmologies.}
\label{t_param}
\end{table}

On Figs.~\ref{ccopt} and \ref{ccinf}, we show multi--wavelength 
counts obtained for the SCDM, $\Lambda$CDM, and OCDM cosmologies defined in 
table~\ref{t_param}. From these figures, one clearly sees that, 
in agreement with Heyl et al. \cite*{HCFN95} and Somerville \&
Primack \cite*{SP00}, the UV to NIR counts are 
relatively insensitive to changes in the cosmological parameters.
This relatively ``stable'' behaviour extends to the FIR range. 
In sharp contrast, the differences between 
the predictions of the various cosmological models are spectacular in the 
submm range : at 850 $\mu$m and for the flux level of 10 mJy, 
the model predicts $\sim 100$ 
times more sources (or $\sim 10$ times brighter sources) in the OCDM 
cosmology than in the SCDM. An interesting trend also comes out of these 
figures: with the quiescent model, any low matter--density universe does a 
better job at matching the ISOPHOT counts at 175 $\mu$m and the SCUBA 
counts at 850 $\mu$m than a critical one. Our OCDM is even able to fit the 
submm counts ``naturally'', without any additional ingredient. 

The influence of cosmology on the faint counts is produced by the complicated 
combination of several effects. For instance, for lower values of the density 
parameter $\Omega_0$, either with zero cosmological constant, or with zero 
curvature, we have the following changes:
\begin{enumerate}
\item{For halos of a given mass, the collapse occurs earlier on an average
(see Fig~\ref{pk_rate});}
\item{The halo number density is lower (see Fig~\ref{pk_rate});}
\item{With a fixed value of $\Omega_B$, the baryon fraction is higher, 
so that there is, on an average, more fuel for star formation per halo of a 
given mass;}
\item{Volume elements are larger;}
\item{Luminosity distances are larger;}
\item{The time versus redshift relationship changes, and the amount of 
evolution undergone by the sources at any redshift with respect to $z=0$ 
is larger.}
\end{enumerate}

These six effects act in different ways. If they are taken separately, 
points 3, 4, and 6 increase the 
slope of the counts whereas points 1, 2, and 5 decrease the slope.  
In addition to this, there is 
the effect of the $k$-correction. In the optical, and NIR, the 
$k$--correction is positive, and cancels out partly what is occurring at 
high redshift in such a way that the faint counts are weakly sensitive to 
cosmology. At optical wavelengths, the net effect is that faint 
counts in low 
matter--density universes are below those in the SCDM. This conclusion is 
opposite to the predictions of phenomenological models based on backward 
evolution of the local luminosity function, under the assumption of 
monolithic collapse and pure luminosity evolution (see e.g. \cite{GRV90}).
The latter models predict that the OCDM is over the SCDM.
The origin of this discrepancy is that the phenomenological models do not take 
points 1, 2 and 3 into account. 

This weak sensitivity extends to the mid--IR and FIR, but the behaviour of the 
submm counts is very different. 
In contrast with the optical, NIR, mid--IR and FIR, the $k$--correction is 
negative at submm wavelengths (see Fig 17 of Paper I for an illustration), 
and enhances the effects of cosmology and evolution at high redshift.
The dominant effects are the earlier collapse (see Fig~\ref{pk_rate})
and larger volume available at high redshift in low matter--density universes.

Such an effect is particularly marked on the redshift distribution of the 
sources given in fig~\ref{rdec}. These predictions are compared with the 
faintest redshift survey in the $I$ band (the CFRS, 
Lilly et al. \cite*{LTHCL95}, Crampton et al. \cite*{CLLH95}),
with $I_{AB}<22.5$, and the 
North Ecliptic Pole Region (NEPR) survey from {\sc iras} at 60 $\mu$m, with 
$60 \le S_{60} \le 150$ mJy \cite{AHHSW96}. The CFRS survey 
is correctly reproduced by the 
quiescent model, whatever the cosmology, though all quiescent models 
seem to overpredict 
low--luminosity objects and produce a peak at too low a redshift 
with respect to the data ($\sim 0.3$ 
instead of $\sim 0.5$). This is clearly due to the overproduction of 
low--luminosity objects in the luminosity function. There is not much 
sensitivity to cosmology.  At 60 $\mu$m, the various cosmologies predict 
different redshift distributions. The SDCM peaks at low redshift, without 
any high--redshift tail, as anticipated by GHBM. The $\Lambda$CDM and the
OCDM peak at higher redshift, with broader distributions.
The OCDM already seems to overpredict high--$z$ galaxies in the NEPR at 
60 $\mu$m. Finally, the sensitivity of 
the redshift distributions to cosmology is spectacular in the submm range.
The wavelengths 175 $\mu$m and 850 $\mu$m and the 
flux cuts at 100 and 2 mJy respectively correspond to on--going redshift 
surveys of the ISOPHOT and SCUBA sources. There are no firm results for 
these surveys because of identification problems (Downes et al. \cite*{DN+al99},
Smail et al. \cite*{SIKCBBOM99}), and we prefer not to plot data. 

However, our quiescent models do not contain mergers and therefore do not 
account for the massive ULIRGs seen by ISOPHOT and SCUBA. This has to be 
taken into account in the model, though a low matter--density universe
lessens noticeably the importance of the contribution of ULIRGs to the
cosmic FIR luminosity. We now try to assess the impact of such
mechanisms on our results phenomenologically.

\section{Sensitivity of faint counts to the star formation history}

Our simple SAM is not able to compute either the merging history of halos, 
or of the galaxies they host. However, we know that locally there is a
tight correlation between major mergers on one side, LIRGs and ULIRGs 
on the other: at least 95 \% of them
are currently undergoing major mergers (see for instance Sanders \&
Mirabel \cite*{SM96}). It also seems fairly safe to assume that ISOPHOT and 
SCUBA sources are the high-redshift
counterparts of such mergers. As a matter of fact, one could sum up the 
qualitative information
from currently available datasets as follows. First, 
the objects seen by SCUBA have to be either very massive, or very efficient to 
extract energy from the gas,
simply because their bolometric luminosity is larger than $10^{12} L_\odot$. 
Second, they have to be highly 
extinguished because most of this luminosity is emitted in the IR/submm.
Third, for such numerous bright sources not to have been detected
in the IRAS NEPR redshift survey at 60 $\mu$m, they have to be located in 
majority at redshifts greater than about $\sim 1.5$, which 
seems to be the case for some of the SCUBA sources \cite{BCSIBK99}.

\begin{figure}[htbp]
\resizebox{9cm}{!}{\includegraphics{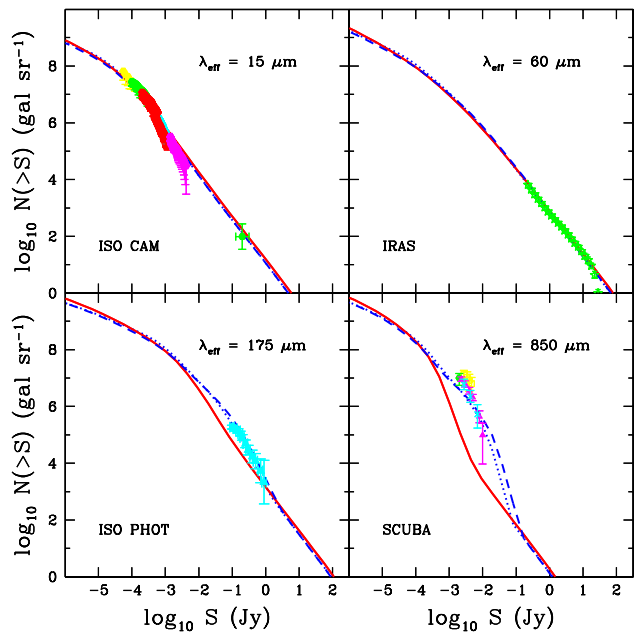}}
\hfill
\caption{IR counts for the SCDM ``quiescent'' model (solid lines), 
the ``burst'' model (dots), and a still more efficient model to extract 
luminosity from the gas (dashes). The luminosities (resp. number densities)
of ULIRGs are multiplied (resp. divided) by a factor 2 in the efficient 
model as compared to the ``burst'' model.}
\label{ix2}
\end{figure}

In light of these observational facts, and as in GBHM, we   
define an {\em ad-hoc} ``starburst'' model, simply by pushing  
the limits of our quiescent models (SCDM, OCDM, or $\Lambda$CDM), 
still powering the sources with star formation according to a Salpeter IMF.  
This consists merely in transforming a fraction of high--redshift
quiescent objects into ULIRGs, while keeping all the parameters of the model 
fixed. The obvious interest of such an exercise is to assess whether
one is able to reproduce the SCUBA source counts, along with preserving the 
quality of the fits of the optical counts 
used to calibrate the quiescent model, in the various 
cosmologies. We hereafter focus on the SCDM cosmology, for which the 
``quiescent'' mode of star formation is unable to reproduce the submm counts. 
 
\begin{figure}[htbp]
\resizebox{9cm}{!}{\includegraphics{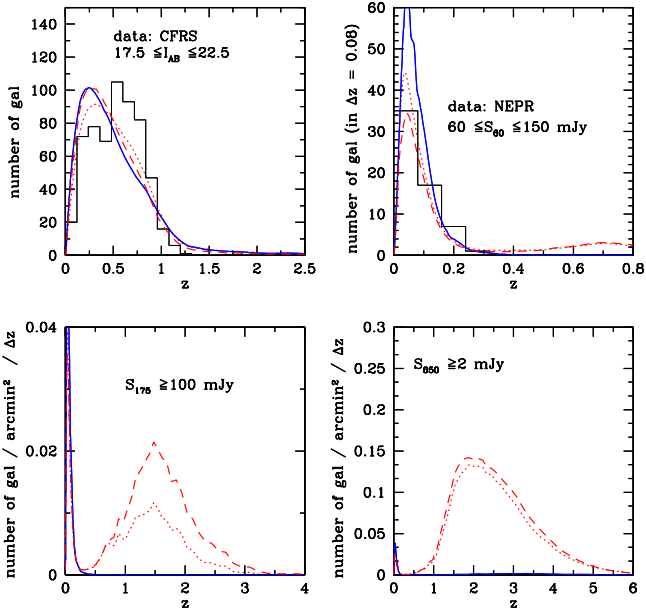}}
\hfill
\caption {Multi--wavelength redshift distributions for the SCDM ``quiescent'' 
model (solid lines), the ``burst'' model (dots), and a still more efficient 
model to extract luminosity from the gas (dashes), as in Fig~\ref{ix2}.}
\label{rx2}
\end{figure}

In order to build such an {\em ad-hoc} model, we use the reasonable recipe 
that follows:
\begin{enumerate}
\item A fraction of objects with halo masses 
larger than $10^{12} M_\odot$ goes through a ULIRG phase
when their host halos collapse; their SFRs and optical 
depths are typically two orders of magnitude higher than those of the 
$z=0$ Milky Way ({\em e.g.} Rigopoulou et al. \cite*{RLRR96}). We tune our 
$\beta$ and $f_c$ parameters to obtain such properties for the 
heavily--extinguished burst mode of star formation. Typically, we take 
$\beta=0.5$ and $f_c=0.5$ for the starbursts. 
\item These ULIRGs are mainly located at $z>1.5$, which we enforce 
by requiring that their fraction evolves proportionally to 
the squared density, {\em i.e.} as $(1+z_{coll})^6$.
\item Their number density at redshift 0 
is consistent with the {\sc iras} luminosity function of \cite{SN91}. 
\end{enumerate}

As a result of this phenomenological recipe, a typical halo of 
mass $10^{12} M_\odot$, with reduced spin parameter $\lambda \simeq 0.04$, 
that collapses at redshift $\simeq 3$, hosts by redshift 
$ \simeq 2.8$ (180 Myr after the starburst was triggered) 
a ULIRG of size 1 kpc that has consumed 98 \% of its $10^{11}
M_\odot$ of cold gas initially present. The star formation rate averaged over 
this period is $\sim 540~M_\odot$ yr$^{-1}$.
The starburst galaxy has a typical column density of 
about $7 \times 10^{22} \mathrm{at~cm^{-2}}$, and a 
metallicity of 0.03, yielding 
a face--on optical depth in the $B$ band of 128. Its absolute 
$B$ magnitude and bolometric IR luminosity (between 3 and 1000 $\mu$m) reach
$M_B = -19.58$ and $2 \times 10^{12} L_\odot$ respectively. 

\begin{figure}[htbp]
\resizebox{9cm}{!}{\includegraphics{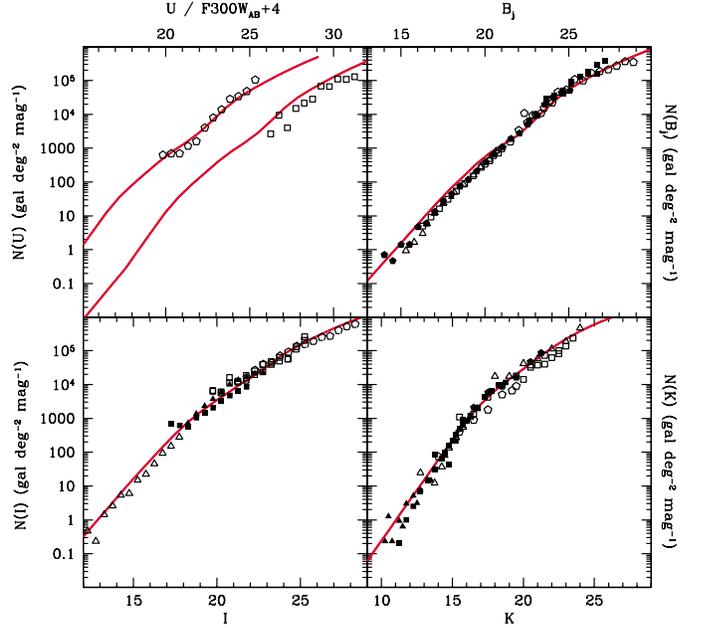}}
\hfill 
\caption{UV/near-IR counts for the fiducial model (solid line).}
\label{fido}

\end{figure}

\begin{figure}[htbp]
\resizebox{9cm}{!}{\includegraphics{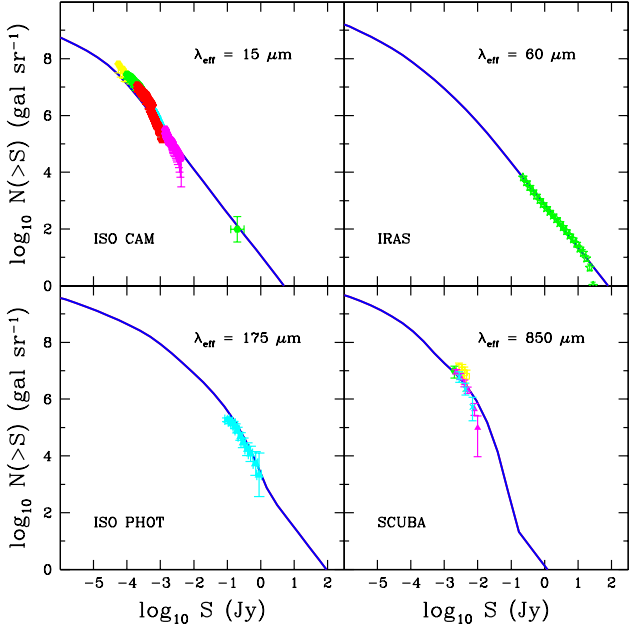}}
\hfill
\caption{Mid--IR/submm counts for the fiducial model (solid line).}
\label{fidi}

\end{figure}

\begin{figure}[htbp]
\resizebox{9cm}{!}{\includegraphics{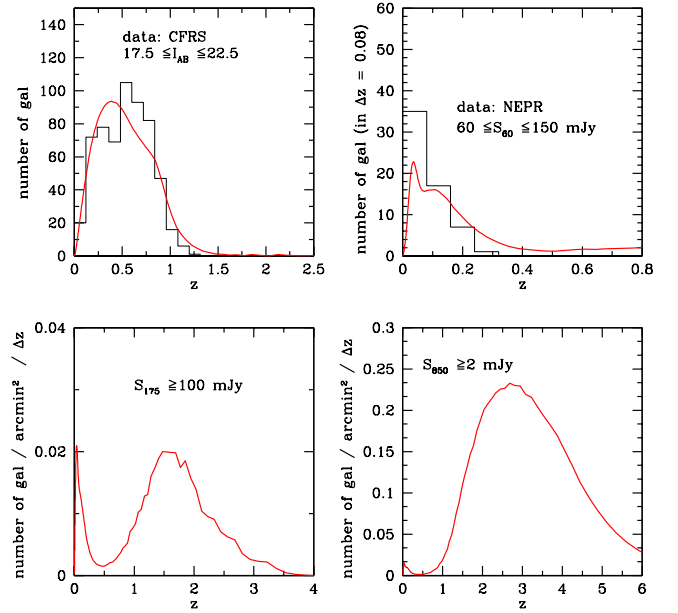}}
\hfill
\caption{Multi--wavelength redshift distributions
for the fiducial model (solid line).}
\label{fidr}
\end{figure}

\begin{figure}[htbp]
\resizebox{9cm}{!}{\includegraphics{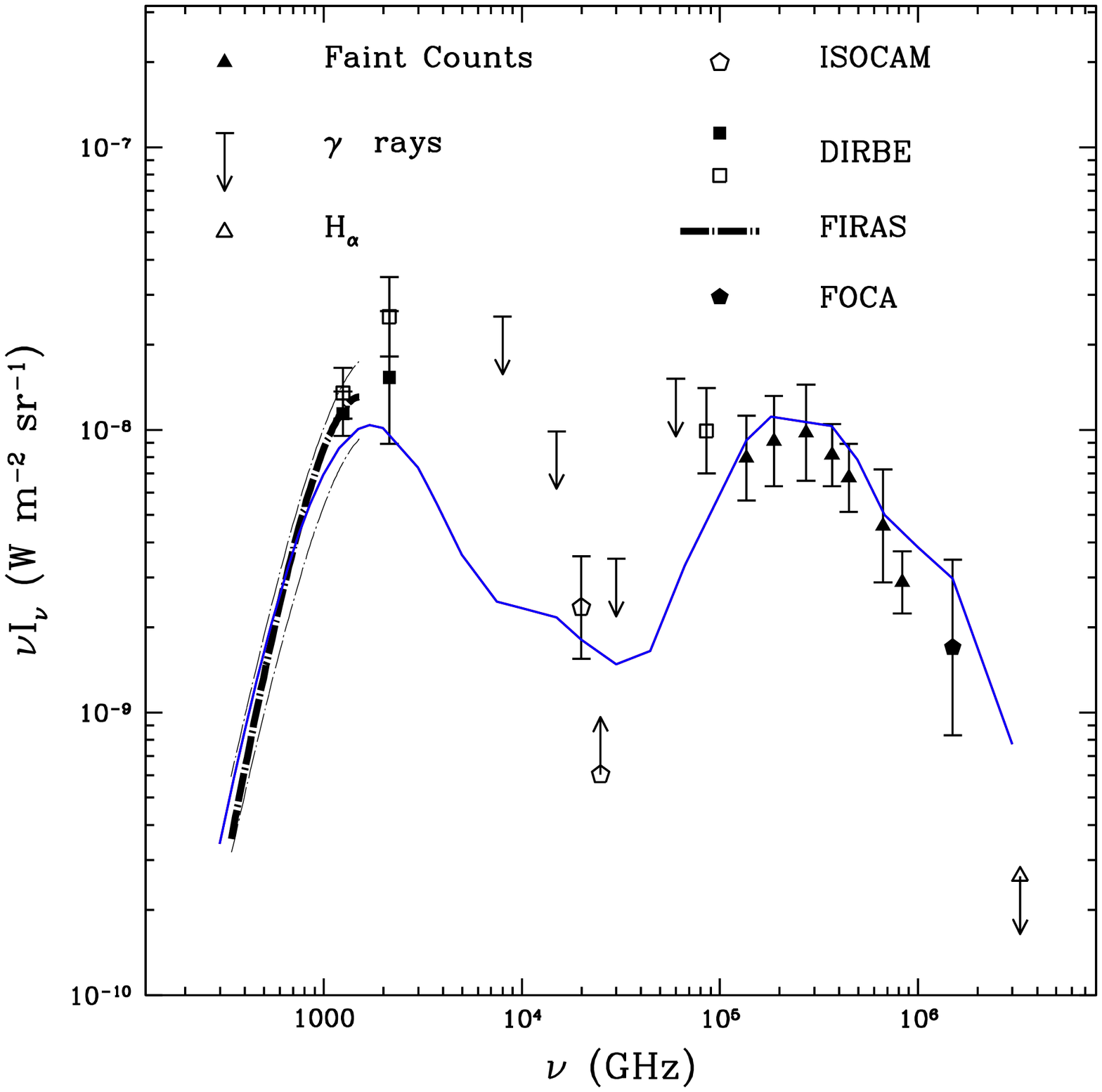}}
\hfill
\caption{Diffuse Background light for the
fiducial model (solid line).}
\label{fidb}
\end{figure}

Of course, such a model is quite drastic, but once again, 
it should be considered as the necessary extension of the quiescent models 
to produce the correct amount of FIR/submm luminosity.
The interesting result is that such a SCDM model 
in which {\em all} massive objects that form at redshifts higher 
than 1.5 are ULIRGs produces almost enough IR/submm luminosity to match the 
ISOPHOT and SCUBA counts, as can be 
seen in Fig~\ref{ix2}. This is also the typical luminosity one can 
extract from star formation with a Salpeter IMF without ruining the UV/IR 
calibration of the counts. For instance, decreasing the mass above which 
the ULIRG phenomenon occurs by an order of magnitude strongly decreases 
the optical counts.

We also have to examine the possibility that a more efficient 
mechanism powers these sources, for instance a top--heavy IMF, with all
the energy available through stellar nucleosynthesis being 
reprocessed in the IR/submm. The main features of such a model 
have been discussed in GHBM who take this solution to accommodate submm counts 
easily in an SCDM cosmology. We refer the reader to that paper for details. 
To test this possibility, we simply take our burst model and multiply
the luminosity output of each ULIRG in the infrared per unit mass, 
$L_\mathrm{IR}/M$, by a factor 2, while lowering the number of ULIRGs in 
the model by 2. This is to say, we trade the number of sources for more 
luminosity per source. Fig~\ref{ix2} and Fig~\ref{rx2} show that the 
influence of such a redistribution on the counts is weak. Of course, any 
combination of luminosity and number density of ULIRGs is possible.

In light of the previous work, and bearing in mind that we want to describe
multi--wavelength galaxy counts, we can define a ``best guess'' model within
a given cosmological model. We hereafter retain the $\Lambda$CDM model as a
typical example, since the optical and submm counts with the 
$\Lambda$CDM model and the 
quiescent mode of star formation only are intermediate between the SCDM and 
OCDM. We take $h=0.7$, $\Omega_0=0.3$, $\lambda_0=0.7$, 
$\Omega_B=0.015 h^{-2}$, and $\sigma_{8h^{-1}} = 1$. In terms of our 
astrophysical parameters, we keep the standard value $\epsilon_{SN} = 0.03$, 
and we take $\beta = 50$, $f_c = 6$ for the quiescent galaxies, and 
$\beta=0.5$, $f_c=0.5$ for the starbursts. One ULIRG dwells 
in each halo that is more massive than $10^{12} M_\odot$ and collapses before
redshift 1.5. This ULIRG population evolves as the density 
squared at lower $z$, so that at redshift 0, its number density is 
about $10^{-7}$ Mpc$^{-3}$, corresponding to only one ULIRG for 2500 halos 
$\geq 10^{12} M_\odot$. The predictions for the faint counts 
are given in Figs~\ref{fido}, and \ref{fidi}. The model provides a good fit of 
the faint 
counts at optical wavelengths (though the bright counts are slightly 
overestimated). The quality of the fit nicely compares with other faint 
counts obtained from SAMs (e.g Kauffmann et al. \cite*{KGW94}). The ISOCAM 15 $\mu$m data and 
{\sc iras} 60 $\mu$m data are also fairly reproduced, though the observed 
slope of the 15 $\mu$m counts seem to be slightly steeper than the model. 
The fit of the submm counts is also very satisfactory. 

The redshift distributions 
are given in Fig~\ref{fidr}. The CFRS predictions now peak almost at the 
correct redshift. The NEPR predictions still exhibit a high--redshift tail
as in GHBM, in contrast with the data, but the level is much lower than 
in GHBM. We recall that the NEPR sample is polluted by a supercluster in the 
first redshift bin. Moreover, a recent follow--up of this sample 
with ISOCAM at 15 $\mu$m seems to show that some of the sources 
are multiple and that the optical identifications might be ambiguous in 
these cases \cite{ACMZF00}. The relative levels of the two peaks in the 
redshift distribution at 175 $\mu$m are sensitive to the flux cut--off. Most 
of the sources in the redshift distribution for the SCUBA deep 
surveys at 850 $\mu$m are predicted to be at $z>1$, but the comparison with 
data is still difficult because of identification uncertainties 
(see e.g. Barger et al. \cite*{BCSIBK99} corrected after Smail et al. \cite*{SIKCBBOM99}).

Finally, Fig~\ref{fidb} shows the Cosmic Background obtained by integrating 
the faint counts, and compare the predictions with current data in the 
optical, IR and submm. Whereas introducing ULIRGs in an {\em ad--hoc} way 
into our simple models suffices to reproduce the Cosmic IR Background 
and the submm counts at 850 $\mu$m, it falls marginally short of 
getting the required 
diffuse background flux at 140 and 240 $\mu$m, 
though it reproduces the ISOPHOT counts brighter than 100 mJy at 175 $\mu$m.
These galaxies contribute only 10 \% of the background. So this discrepancy 
may be due only to the fact that the 175 $\mu$m counts below 100 mJy are 
much steeper than our predictions. The model is too low by a 
factor of 2 with respect to the points corrected for warm galactic 
dust by Lagache et al. \cite*{LABDP99}, which are themselves a factor of 
1.5 below the points without such a correction by Hauser et al. \cite*{H+al98}. 
The difficulty to fit the points might indicate that this correction 
is still underestimated. Finally,
one should also be aware that a contribution of intergalactic dust (with a 
grey extinction curve) to the background light is also possible \cite{AH99}.
Adding these extra components might help reconcile models and observations.

We conclude from these figures that this fiducial model gives a satisfactory 
estimate of the luminosity budget of galaxies, and allows us to interpolate
or extrapolate the observed faint counts to other wavelengths 
and fainter flux levels.

\section{Discussion and conclusions}
 
In this paper, we have proposed a first implementation of the set 
of {\sc stardust} synthetic spectra into a SAM of galaxy 
formation. Although our model is quite simple, and cannot properly handle 
the merging history trees of halos and galaxies, we have shown that the 
implementation of {\sc stardust} is quite straightforward. This 
implementation can be 
easily achieved in more sophisticated SAMs.

We have not explored a large set of statistical properties in this paper, 
but have only illustrated the ability of this approach to reproduce 
the optical/IR/submm luminosity budget by producing predictions of faint 
galaxy counts at UV, visible, NIR, mid--IR, 
FIR, and submm wavelengths. 
As in GHBM, we have defined a quiescent mode of star formation which corresponds to 
the ``natural'' values of these astrophysical parameters as 
they are suggested by local observations of disks (for $\beta$ and $f_c$), 
or by numerical simulations (for $\epsilon_{SN}$). 

We have then studied the influence 
of the cosmological parameters $\Omega_0$ and $\Lambda$ on the faint counts.
In SAMs, the cosmological parameters influence the counts at various stages
of the computation, through halo collapse as well as through the relationship 
of the cosmic times, luminosity distances and volume elements to redshifts. 
These quantities intervene in the computation of the counts in different 
ways, and their effect is dimmed or enhanced by the $k$--corrections.  
The net result on the optical and NIR counts,
with the influence of a positive $k$--correction, is a rather weak
sensitivity to the values of the cosmological parameters. 
In contrast, because of the negative $k$--correction that enhances
what happens at high redshift, the submm counts show a strong sensitivity to 
cosmology. 

We know that some of the SCUBA sources detected at 850 $\mu$m are the 
high--redshift counterparts of local LIRGs and ULIRGs. These objects harbour
heavily--extinguished starbursts due to gas inflows triggered by 
interaction/merging. Our SAM is unable to address this process, and, as in 
GHBM, we chose to implement a heavily--extinguished starburst mode of star 
formation, by increasing the number
fraction of massive objects that undergo this stage, as the squared density.
Of course, the LIRGs and ULIRGs can also be powered by a top--heavy IMF. 
We have shown the strong sensitivity 
of the submm counts to the details of the ULIRG scenario, because of the 
negative $k$--correction, contrasting it with the weak influence on 
the optical counts. 

We have also produced redshift distributions at various wavelengths and 
flux cuts. Here again, the sensitivity of the results to cosmology and 
evolution is dramatic at submm wavelengths. We got a fair fit of the CFRS 
redshift distribution in the $I$ band, and the redshift distribution 
predicted for the NEPR survey looks much better than in GHBM. 
At submm wavelengths, we predict
a double--peaked distribution with nearby (mostly quiescent) sources and 
distant (mostly starburst) sources. The relative weight of the two broad
peaks is sensitive to cosmology and evolution. Sufficient statistics in an
observational sample could in principle help disentangling these effects.
Unfortunately, the identification process of 
the submm sources is difficult, either in ISOPHOT or in SCUBA samples,
and we might have to wait for multi--wavelength observations with forthcoming 
satellites such as {\sc sirtf} and {\sc first} to solve this issue.
In this context, our predicted counts are also a useful ingredient to 
analyse current data, and to prepare
observational strategies with these satellites. In this purview, 
we have proposed a new ``fiducial model'' in a $\Lambda$CDM cosmology 
which can take the place of the model ``E'' in GHBM, and which is available 
upon request. 

Merging--triggered starbursts and subsequent bulge formation are 
key--processes in the paradigm of hierarchical galaxy formation. 
It turns out that these processes are particularly apparent at IR/submm 
wavelengths, and almost invisible at optical wavelengths. 
So multi--wavelength observations are required to constrain the history of 
galaxy formation in the quiescent and starburst modes. The complete merging 
history of galaxies has to be followed in detail, especially 
during the short periods of interaction and merging which produce 
IR luminous sources.
A hybrid method of galaxy formation using high--resolution N--body
simulations to plant semi--analytic galaxies is being 
developed and will be used to quantify the importance of merging processes
(Hatton et al. \cite*{HNDBGSV00} and following papers). In this context,
the reader is invited to notice the following point:
any change in the luminosities and number densities of the ULIRGs 
results into spectacular changes in the submm counts, 
because of their super--Euclidean regime.  
This sort of ``instabilities'' in the behaviour of the faint counts is 
going to be very useful to constrain the luminosity and 
duration of the starbursts in more refined models.

Throughout this paper, we have assumed that dust heating is powered by 
star formation. The other possible engine is the presence of 
heavily--extinguished AGNs located at the centers of ULIRGs. In local samples, 
heating is dominated by starbursts, except in the most luminous galaxies 
(Genzel et al. \cite*{G+al98}, Lutz et al. \cite*{LSRMG98}). 
So far, we do not know how this 
situation evolves with redshift. The redshift distribution of the SCUBA 
sources seems to peak around redshift 2.5 where the
observed redshift distribution of quasar activity may also peak \cite{P95}.
Perhaps a sophisticated combination of AGNs and starbursts with a top--heavy 
IMF is responsible for the evolution seen by SCUBA, and we signal
that attempts to include both types of ingredients are already discussed in 
the literature (e.g. Blain et al. \cite*{BJSLKI99}).
However, properly disentangling all these components
definitely requires a more sophisticated model than the one presented here.
We therefore defer this issue to a later study, but remark that 
AGNs can be implemented self--consistently within the framework of the hybrid 
method \cite{KH00}, so that one could hope to set quantitative limits on 
their respective contributions to the submm fluxes. 

The situation is indeed as complex from the theoretical as well as the
observational points of view. 
That is why this simple modelling should be considered as an exploratory step
to probe the complex issues involved in multi--wavelength counts, and to 
design a more satisfying approach of the problem.

\begin{acknowledgements}
We are pleased to thank David Elbaz for communicating pertinent comments on an 
early version of this paper as well as for providing us with his data in
electronic form. We also acknowledge fruitful discussion with 
Herv\'e Aussel, Fran\c{c}ois R. Bouchet, Guilaine Lagache, and 
Jean--Loup Puget. 
\end{acknowledgements}

\nocite{HPCCBSS97}
\nocite{W+al96}
\nocite{ALMMMBK97}
\nocite{BD97}
\nocite{GSCF96}
\nocite{MSFGR96} 
\nocite{WDF95}
\nocite{SHYC95}
\nocite{LTHCL95} 
\nocite{MDGSPY97} 
\nocite{D+al95}
\nocite{K+al98}
\nocite{EC+al99}

\appendix

\section{Dark matter halos in any cosmology}

We suppose that perturbations of the matter density field when the
universe becomes matter dominated
are completely characterised by their power spectrum $P(k)$:
\begin{eqnarray}
P(k) \propto k^n T^2(k) ,
\end{eqnarray} 
where $T(k)$ is the transfer function (see fit given in Appendix G of
Bardeen et al. \cite*{BBKS86}). We further assume a post-inflation
Harrison--Zel'dovich power spectrum ($n = 1$) for these perturbations, and 
take the shape parameter $\Gamma$ used in the computation of $T(k)$ to  
be \cite{S95}: 
\begin{eqnarray}
\Gamma = \Omega_0 h \exp \left[ -\Omega_B 
( 1+ \frac{\sqrt{h/0.5}}{\Omega_0}) \right] ,
\end{eqnarray}
where $\Omega_0$ is the current matter density (in critical density units),
$\Omega_B$ is the baryon density, and $h = H_0 / (100~\mathrm{km/s/Mpc})$ is 
the reduced Hubble constant.  

In the linear regime, the equation of motion is solved for 
the expansion factor, $a$, and the solutions for the 
growth of the density contrast $\delta$ (see {\em e.g.} Peebles \cite*{P80})
are derived. There are two such solutions \cite{H77}, which form a complete
set \cite{Z65} and read:
\begin{eqnarray}
D_d[z] &=& H_0 \left\{ \Omega_0(1+z)^3+(1-\Omega_0-\lambda_0) 
    (1+z)^2+\lambda_0 \right\} ^\frac{1}{2} \\
D_g[z] &=& D_d[z] \, a_0^2 {\int^{\infty}_{z} \frac{(1+x)} 
    {D_d^3 [x]} \mathrm{d} x } ,
\end{eqnarray} 
where the subscripts $d$ and $g$ respectively stand for the decaying 
and growing modes, and $\lambda_0 = \Lambda/(3H_0^2)$ is the reduced 
cosmological constant.
If one further assumes that the initial peculiar velocity of the perturbation
is zero ({\em i.e.} that the perturbation simply moves along with the
expanding universe), the 
density contrast $\delta [z]$ in the linear regime grows as: 
\begin{eqnarray}
\delta [z] & = & \left\{\frac{3}{2}\Omega_0
         (1+z_i)+1-\Omega_0-\lambda_0 \right\} \frac{H_0^2}{a_0^2}
         \, D_g[z] \, \delta_i ,
\label{del}
\end{eqnarray}
where the subscript $i$ stands for the initial quantities.

In the non--linear regime, one considers an isolated
spherical perturbation of radius $r_i$, 
at time $t_i$, which has a uniform
overdensity $\delta_i \equiv (\rho_p[t_i] - \rho_b[t_i])/ \rho_b[t_i]$
with respect to the background ($\delta_i \ll 1 $), and encloses a mass 
$ M = 4/3 \pi r_i^3 \rho_b [t_i] (1+\delta_i) $. We assume that the 
perturbation is bound, and that its peculiar velocity is nil.
The equation of motion is integrated to obtain the time 
$t_{m}$ at which the perturbation reaches its maximum expansion radius 
$r_{m}$:
\begin{eqnarray}
t_{m} = t_i + \frac{1}{H_0} {\int_1^{r_{m}/r_i}
\frac{x^{1/2}}{D_p[x]^{1/2}} \, \mathrm{d}x } ,
\end{eqnarray}
where 
\begin{eqnarray}
D_p[x]&=& \lambda_0 \, x^3 + \{ (1 - \Omega_0 - \lambda_0) (1+z_i)^2  \\ 
  &-& \Omega_0 \delta_i (1+z_i)^3  \} \, x 
  + \Omega_0 (1 + \delta_i) (1+z_i)^3 ,
\end{eqnarray}
and $r_{m}/r_i$ is the first real root $ > 1$ of the cubic
equation ({\em c.f.} Richstone et al. \cite*{RLT92}): 
\begin{eqnarray}
D_p[x] = 0 .
\end{eqnarray}
After it has reached this maximum radius at time $t_{m}$, 
the perturbation, by symmetry, collapses on a time scale
$t_{coll}= 2 t_{m} - t_i$. Thus, with the previous equations, 
the critical density contrast $\delta_0$ linearly extrapolated till 
today (eq.~\ref{del}) is explicitly related to the collapse redshift 
$z_{coll}$ of the perturbation for any cosmology, just by computing 
the redshift to which the collapse time 
of a perturbation with overdensity $\delta_i$ corresponds 
in the {\em unperturbed} universe (provided $t_{coll}$ is smaller
than the age of the universe):
\begin{eqnarray}
t_{coll} \equiv {\int_{z_{coll}}^{\infty} \frac{1}{(1+x) 
D_d[x]} \, \mathrm{d}x } .
\end{eqnarray}

\begin{figure}[htbp]

\resizebox{9cm}{!}{\includegraphics{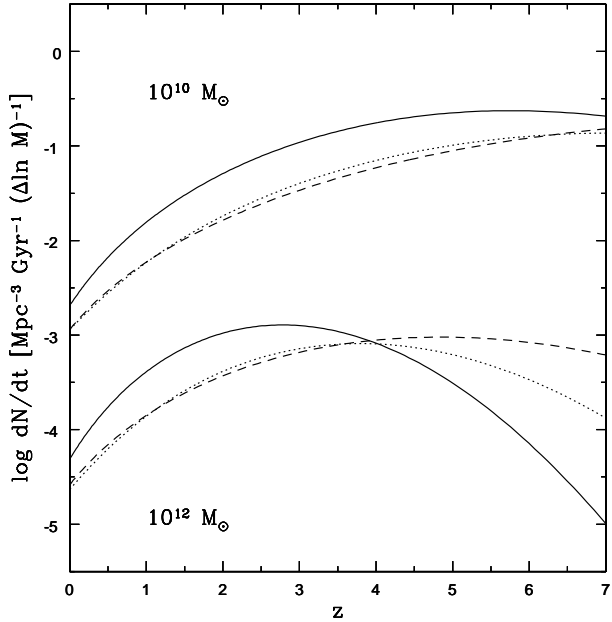}}
\hfill
 \caption  {Evolution of the formation rate 
of dark matter halos (masses indicated 
on the figure) for three different cosmologies: 
SCDM (solid line), OCDM (dashes), and
$\Lambda$CDM (dots). The parameters used for these models 
are given in table \ref{t_param}.
}
\label{pk_rate}

\end{figure}

If the resulting virialized perturbation can be approximated by
a singular isothermal sphere truncated at virial radius 
$ r_{vir}$, then $ r_{vir}/r_{m} $ is the 
solution of the following cubic equation (see Devriendt \cite*{Dphd99}):
\begin{eqnarray}
\frac{4 \lambda_0 H_0^2 r_{m}^3} {3 G M} x^3 -
  (\frac{12}{5} + \frac{6 \lambda_0
  H_0^2 r_{m}^3}{5 G M}) x + 1 = 0.
\end{eqnarray}
Finally, the velocity of a test particle moving on a circular
orbit in the isothermal sphere reads:
\begin{eqnarray}
V_c = \sqrt{2} \sigma = \sqrt {\frac{G M}{r_{vir}} - \frac{2\lambda_0}{9} H_0^2 r_{vir}^2} .
\label{vc}
\end{eqnarray}

Implementing these results into the peaks formalism described in GHBM enables
one to derive formation rates for dark matter halos as a function of redshift.
Fig~\ref{pk_rate} illustrates this halo formation rate for three typical 
cosmologies gathered in table~\ref{t_param}.

\end{document}